# Comparing of Term Clustering Frameworks for Modular Ontology Learning

Ziwei XU[1], Mounira HARZALLAH[1] and Fabrice GUILLET[1]

[1]*LS2N, Polytech'Nantes ,Rue Christian Pauc ,44300, Nantes, France*
*{Ziwei.Xu, Mounira.Harzallah, Fabrice.Guillet}@univ-nantes.fr*

Abstract: This paper aims to use term clustering to build a modular ontology according to core ontology from domain-specific text. The acquisition of semantic knowledge focuses on noun phrase appearing with the same syntactic roles in relation to a verb or its preposition combination in a sentence. The construction of this co-occurrence matrix from context helps to build feature space of noun phrases, which is then transformed to several encoding representations including feature selection and dimensionality reduction. In addition, the content has also been presented with the construction of word vectors. These representations are clustered respectively with K-Means and Affinity Propagation (AP) methods, which differentiate into the term clustering frameworks. Due to the randomness of K-Means, iteration efforts are adopted to find the optimal parameter. The frameworks are evaluated extensively where AP shows dominant effectiveness for co-occurred terms and NMF encoding technique is salient by its promising facilities in feature compression.

## 1 INTRODUCTION

Ontology building is a complex process composed of several tasks: term or concept acquisition, concept formation, taxonomy definition, ad-hoc relation definition, axiom definition, etc. (Fernández-López et al., 1997). The ever-increasing access to textual sources has motivated the development of ontology learning approaches based on techniques of different fields, like natural language processing, data mining and machine learning. Many works are focused on the taxonomy definition and more especially on the hypernym relation extraction. A term t1 is a hypernym of a term t2 if the former categorizes the later. This relation is also known as a terminological « is-a » relation. For its extraction from texts, several approaches based on Harris' distributional hypothesis are proposed. This hypothesis states that words/terms in the same context can have similar meanings (Harris, 1954). Then, each term can be represented as *a vector of contexts*, forming a matrix of co-occurrence or colocation (i.e. co-occurrence of the second order). Based on the semantic similarity in a vector space, non-supervised methods are applied for term clustering. Each cluster is expected to include semantically similar terms (i.e. synonyms or related by the hypernym relation) or semantically connected terms.

However, obtained clusters are not necessarily relevant for the ontology to build. Moreover, these approaches may have a poor performance due to the sparsity of the co-occurrence matrix (Buitelaar et al., 2004). Dimensionality reduction becomes a crucial issue. It can be performed by feature selection. In statistical stage, feature selection could be achieved by the frequency of terms or the weighting of Tf-Idf (term frequency- inverse document frequency).

In our work, we are interested in term clustering according to core ontology in order to build a modular ontology (Kutz and Hois, 2012). A core ontology of a domain is a basic and minimal ontology composed only of the minimal concepts (i.e core concepts) and the principal relations between them that allow defining the other concepts of the domain (Oberle et al., 2006; Burita et al., 2012). This step (i.e. term clustering according to a core ontology) is the first stage towards a taxonomy definition. Indeed, a term of each cluster is expected to be synonym or hyponym of the core concept that corresponds to its cluster. Later, inside of each cluster, other hypernym relations between terms have to be extracted.

In this paper, we analyze and evaluate two frameworks of terms clustering following the processing workflow of Figure 1. We discuss some works dealing with term clustering in section 2. We then describe the resources used for our experiments and preprocessing steps performed in section 3 and expose feature space representation in section 4. Sequently, we discuss the parameters of two clustering techniques, analysis their results and give instructions on future work.

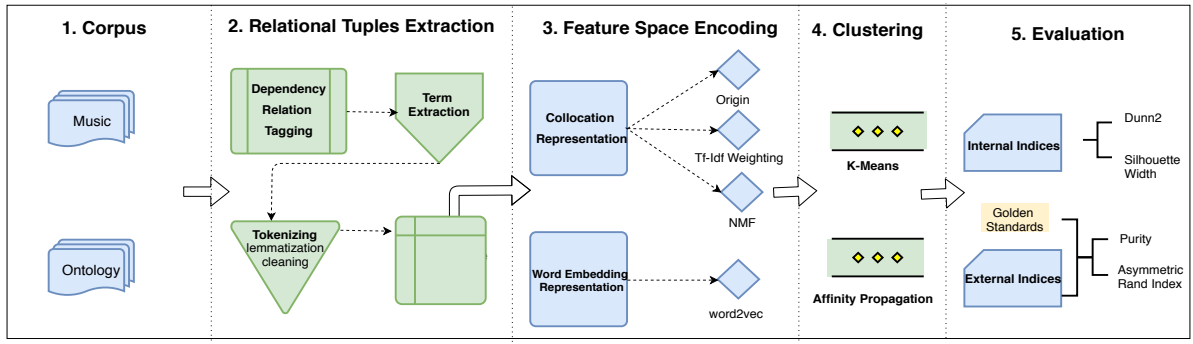

Figure 1 : The Processing Workflow

## 2 RELATED WORK

In the field of knowledge acquisition on feature selection, clauses and their functional equivalent entities are apparent linguistic elements to collect syntagmatic information. Cimiano et al. (2004) describe the local context by extracting triples of nouns, their syntactic roles, and co-occurred verbs. They consider only verbs/object relations, so as to emphasize partial features of terms working as object by conditional probability measure. Similarly, ASIUM (Faure and Nedellec,1998) acquires semantic knowledge from case frames which include the headword of noun phrases, verbs, and their preposition or syntactic roles. For examples, for this sentence "Bart travels by boat", we get

<to travel> <subject> <***Bart***>
<by> <***boat***>

Besides syntactic dependency, one recent work by Gábor et al. (2016) extract co-occurring couples of entities and present their semantic relations with pattern-based representation. To interpret these appearances, terms(entities) are presented by vectors with frequent sequential pattern as components. Then *pattern-based feature space* is constructed for relation discovery. Moreover, according to *Word2vec* (Mikolov et al., 2013), a term is statistically encoded with analogies from its appearance in different context, where the similarity of encoding vectors reflect the semantic relations between terms.

Feature transformation is consistently discussed in order to emphasize characteristics of extracted phrases or sentences. Gábor et al. (2016) proposed to apply PPMI weighting (positive pointwise mutual information) to reduce bias in rare contexts, in which values below 0 are replaced by 0. Tf-Idf (term frequency-inverse document frequency) also contribute to weight terms by their specificity to documents. The computational complexity grows exponentially with the size of the lattice, where *NMF (non-Negative Matrix Factorization)* (Lee and Seung, 1999) is dedicated to solving the dimensionality reduction problem by performing feature compression.

Hierarchical clustering is preferred in terms aggregation, which provides subsumption relation of concepts for ontology learning. Based on that, ASIUM creates conceptual clustering to aggregate clusters for new concepts discovery. Besides, non-hierarchical clustering organizes terms with different relations. *Affinity Propagation* (Frey and Dueck, 2007) gives the link of terms represented by context concepts of message passing between data points.

## 3 RESOURCES AND PREPROCESSING

### 3.1 Resources

For the purpose of term clustering experiments, we choose two corpora about two different domains: music domain and ontology learning domain. For each corpus, we possess a golden standard, which includes a set of extracted terms that are classified manually over the *core concepts* in the domain.

*Music Corpus*, is composed of 100M-word documents, includes Amazon reviews, music biographies and Wikipedia pages about theory and music genres (Competitions.codalab.org, 2018). We deliberately selected 2000 documents from 105,000 whose content includes the great proportion of terms in predefined golden standards. Also, we have *Ontology Learning Corpus* with 16 scientific articles from the journal in the domain of ontology learning. As shown in Table 1, these two corpora are different in terms of domain and the amounts of docs, however, a great contrast could help researchers to figure out whether different strategies have a stable performance of taxonomy discovery.

The aforementioned *core concepts* are predefined for each domain in the golden standard. As Table 2 shown, 190 relevant terms were labeled into 5 classes for *Music golden standard*, while larger terms were labeled into 8 classes for *Ontology learning golden standard*.

Table 1 : Corpus Size and Statistics

| Corpus | Document Size | Sampling | Sentences | Words | Words /documents |
|---|---|---|---|---|---|
| Music | 105,000 | 2,000 | 33,051 | 762,180 | 381 |
| Ontology | 16 | 16(none) | 4901 | 112,628 | 7,040 |

Table 2 : Golden Standard

| Corpus | #core concepts | nb. of terms classified under core concepts | Labels of Core Concepts |
|---|---|---|---|
| Music | 5 | 190 | Album, Musician, Music Genre, Instruments, Performance |
| Ontology | 8 | 742 | Component, Technique, Ontology, Domain, Tool, User, Step, Resource |

## 3.2 Pre-processing

According to the syntactic roles, the skeleton of a sentence is composed by subject and object along with its corresponding verb. In other words, terms with important syntactic roles cover the most descriptive information in a sentence. Thus noun phrases (NP), acting as subject or object, turn to be highlighted in concept extraction, while verbs reflecting the contextual components are used to present the concrete connection between NPs.

From POS parsing stage, syntactic information is extracted in order to identify NPs acting as subject or object and their co-occurred verbs. In our lab, we propose to use spaCyr (Kenneth and Akitaka, 2018) as a parser tool. It decomposes an entire typical syntactic tree, which shows the overwhelming convenience in postprocessing, comparing to other parser tools, such as cleanNLP and coreNLP.

We start with skeleton terms recognition in each sentence. As shown in the top of Figure 2, terms in a sentence are presented with dependency relation, where the shaded terms have been tagged as subject (nsubj), ROOT and object. Subject ('ontowrapper') and Direct Object ('information') point to ROOT ('extract') with the solid line, while Proposition Object ('on-line resource') indirectly points to ROOT ('extract from') with the solid line. For non-skeleton dependency, they are shown in dashed lines. Further, we need to pay attention to the distinction between passive and active sentence. To simplify the composition of couples, it is feasible to record Passive Subject (nsubjpass) as Direct Object (dobj).

Terms extraction is followed in the subpart of Figure 1. With the help of head pointers, presenting with the dashed line, noun phrases (NPs) and verb-preposition combinations (VPCs) are gathered and extracted in compound format. Then, they are cleaned and lemmalised after tokenization. Finally, the couples of ROOT and skeleton terms are tagged with roles of subject or object within sentences and recorded as intermediate data in replacement of raw context.

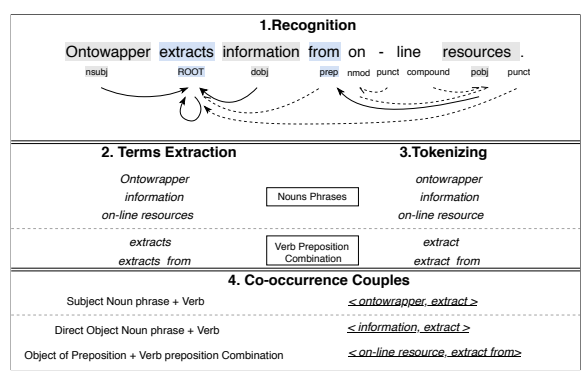

Figure 2 : Instantiated Co-occurrence Couples Extraction

## 4 FEATURE REPRESENTATIONS

We discuss hereinafter two disparate approaches to build basic feature representations. One of the fundamental vector spaces takes advantage of the frequency of NPs and VPCs couples, while another feature representation uses entire context for word embedding. They differ in the range of terms colocation, for which the fundamental method facilitates syntactic roles for co-occurrence couples within a sentence, while the word embedding method takes into account the surrounding context of all appearance places of a term. Additionally, to tackle the sparseness of phrases, dimensionality reduction techniques are employed to condense feature representation.

### 4.1 Co-occurrence Representation

Table 3：Dimensionality Reduction after Threshold

| | #NPs | | | #VPCs | | | Reduction with Frequency | | Reduction with Tf-Idf | |
|---|---|---|---|---|---|---|---|---|---|---|
| | | | | | | | #NPs | #VPCs | #NPs | #VPCs |
| Corpus | subj. | obj. | both | subj. | obj. | both | Threshold $\sigma_1$: Summation of frequency | | Threshold $\sigma_2$: Summation of value | |
| Music | 3138 | 7272 | 1560 | 254 | 3054 | 532 | $\sigma_1>8$ | | $\sigma_2>7$ | |
| | | | | | | | 573 | 660 | 582 | 456 |
| Ontology | 401 | 1643 | 281 | 80 | 889 | 219 | $\sigma_1>3$ | | $\sigma_2>4$ | |
| | | | | | | | 602 | 505 | 563 | 502 |

The aforementioned couples are extracted and transformed into a co-occurrence frequency matrix, where VPCs are features of NPs. Since we notice the big gap of functionality between subject and object, they are organized into separate co-occurrence couples, named subject co-occurrence and object co-occurrence.

Actually, one type of the co-occurrence couples, either subject or object, could only cover segmental linguistic context. It is required to deliberatively combine subject and object co-occurrence couples. In Figure 3, we differentiate NPs and VPCs into 'pure subject', 'pure object' and common part. The common part means NPs and VPCs appear in both subject and object. Entirely, the merged matrix comprises 9 subparts, where the non-existing couples present to be all zero (blank rectangles) and the 'pure couples' (subject or object) present respective frequency in two blue rectangles. Common couples (shaded rectangles), overlapping between subject rectangle and object rectangle, are filled with the cumulate frequency of subject couples and object couples. Positively, as long as subject and object co-occurrence couples join together, the merged matrix theoretically emcompasses complete linguistic information. Hence, it would work as primary representation if encoding techniques are required in the following part.

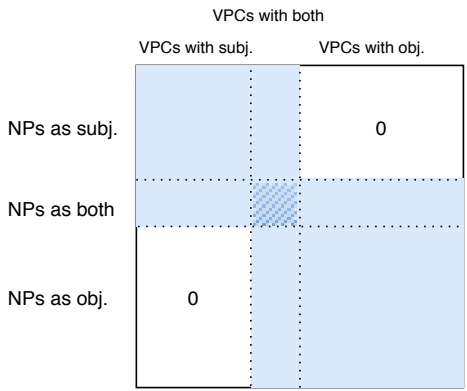

Figure 3: Merged Co-occurrence Matrix

## 4.2 Dimensionality Reduction

The sparseness of the merged co-occurrence matrix becomes a significant issue. Row and column reduction are simultaneously required to decrease the noise effect. In Table 3, a frequency-based threshold $\sigma_1$ could be used to eliminate most common and rare elements. Similarly, Tf-Idf encoding representation provides bi-directional selection respecting to the relevance of NPs to specific VPCs. While NMF encoding is dedicated to reducing feature space.

### 4.2.1 Weighted Co-occurrence

Based on the merged representation, we would like to weight values to decrease the impact of common and rare NPs and differentiate the importance of co-occurrence couples. *Tf-Idf*, is designed with this discriminative purpose. Basically, it extracts the most descriptive terms of documents, which could also extend to weight the most significant NPs to their VPCs, instead of documents. With certain thresholds in rows and columns, only distinguishing NPs and their co-occurred VPCs are kept at last. Thanks to the derivation of Tf-Idf, the strong associated NPs and VPCs are selected according to threshold $\sigma_2$ in Table 3 so that the weighted co-occurrence matrix gets refined with reduced dimensionality.

### 4.2.2 NMF Co-occurrence

Term co-occurrences could be separated into 3 levels according to the identity of words in context (Gamallo and Bordag, 2011). In first-order co-occurrence, terms appear together in identical context. As for two terms are associated by means of second-order co-occurrence, they share at least one-word context and have strong syntactic relations. Besides, terms do not co-occur in context with the same words but between words that can be related through indirect co-occurrences, namely third (higher) order co-occurrence. To manipulate this, *NMF* (Lee and Seung, 1999) is applied to condense isolated VPCs into encoded features. From previous related works, it is reasonable to set the number of features to be 100 during experiments. Then NPs with indirect co-occurrence are signified in the new dense feature space.

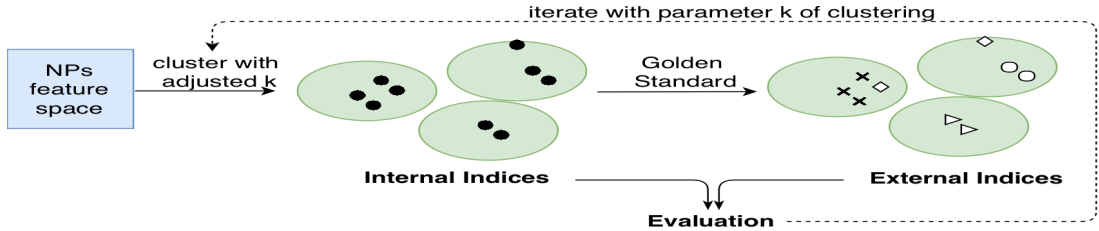

Figure 4: Clustering and Evaluation

### 4.3 Word Embedding Representation

From the distribution of contextual information, it allows building feature vectors that adapt for semantic similarity tasks. Word embedding representation was trained using *word2vec* (Mikolov et al., 2013) algorithm under the skip-gram model. In the local aspect, terms can be represented by vectors of its colocated words within certain window size, called colocating vectors. The sum of colocating vectors around appearance place of a term constitute the context vectors. As for the global aspect, the sum of context vectors at all appearance places of a term gives the construction of word vectors. It integrates all the contextual features of a word and statistically present with encoded similarity. One of the advantages of word2vec is that it embeds techniques to achieve dimension reduction purposes by indicating the required amount of features. To be comparable with NMF encoding technique, the number of features with word2vec is also given by 100.

## 5 CLUSTERING

The most typical clustering technique is *k-means* (Hartigan and Wong, 1979). One of the drawbacks is that k-means is quite sensitive to the initial set of seeds. On the other hand, *affinity propagation* uses graph distance that performs in a 'message passing' way between data points (Frey and Dueck, 2007). It does not need to determine the number of clusters in advance and the centroid of each cluster is specified after calculation, which turns out to be helpful for interpretation. As for distance measurement in clustering, *cosine dissimilarity* is preferred in both techniques because of the high dimension of feature space. As shown in Figure 4, clustering iterates with adjusted parameters k ranging from 2 to 50, indicating the variation of the number of clusters. Each iteration allows for analysis of clustering performance respecting to internal indices and external indices.

## 6 EVALUATION

### 6.1 Evaluation Indices

A large number of indices provide possibilities to assess the clustering quality (Aggarwal and Zhai, 2012). In order to simplify the discrimination process, we select two distinct indices respectively for internal evaluation and external evaluation.

#### 6.1.1 Indices for Internal Evaluation

*Silhouette width* (Rdrr.io., 2018) and adjusted Dunn Index are chosen as indices of internal evaluation. Silhouette method specifies how well each object lies within its cluster.

$$s(i) = \frac{b(i) - a(i)}{\max(a(i), b(i))} \quad (1)$$

In equation 1, $a(i)$ represents average dissimilarity between *i* and all other points of the cluster to which *i* belongs. For all other clusters $d(i, C)$, denotes average dissimilarity of *i* to all observations of C. $b(i)$ is set by the smallest $d(i, C)$ and can be seen as the dissimilarity between *i* and its "neighbor" cluster. A high average silhouette width indicates a good clustering according to features.

*Adjusted Dunn Index* proposed by Pal and Biswas (1997) overcomes the presence of noise comparing to original Dunn Index (Dunn, 1974). In general, they are both dedicated for the identification of "compact and well-separated clusters". Higher values are preferred, which shows a good performance of compactness. Notably, the Dunn Index family does not exhibit any trend with respect to the number of clusters, of which this property is exceedingly welcomed since the number of clusters varies in different iterations.

#### 6.1.2 Indices for External Evaluation

In the case of external evaluation, the indices are slightly different from formers because of the use of a golden standard. According to expected core concept classes, *Purity* and *Asymmetric Rand Index* are representative of clustering quality measurement. To compute *purity*, each cluster firstly is assigned with a label that is most frequent in it, according to the gold standard, then this assignment is calculated by counting the number of correctly assigned elements dividing by all elements. High purity is easy to achieve when the number of clusters is large. A larger amount of clusters may refine the branches of structure in ontology building, however, it incurs complexity to label clusters with core concepts, performing as the first step of ontology learning. Thus we could not use only purity to trade off the quality of clustering against the number of clusters.

The *Asymmetric Rand Index* proposed by Hubert L. and Arabie P. (1985) is also considered, for which it provides the comparison between 2 different sets of labels of same extracted data. It differs from typical Rand Index because it allows for inclusion from a greater number of partitions to relevant partitions. For example, the partitions varied with clustering are always larger than classes of core concepts. The characteristic of inclusion provides a more accurate calculation.

## 6.2 Repetitions with Number of Clusters

To weaken the impact of randomness of clustering, each experiment is repeated 10 times to go through all parameters of k ranging from 2 to 50, so as to get the convincing results with mean values. Thus each index is statistically averaged to be presented for evaluation. For example, as Figure 5 shows, curves of all indices for NP_VPC representations are plotted separately in Ontology Corpus and in Music Learning Corpus. To select the optimal amount of clusters, we attempt to solve the multicriteria optimization problem by finding the first peak of a fluctuating line and assuring a rather higher summation over the entire indices. The dashed lines indicate the final parameter choice for this specific representation. Besides, optimal parameters of the rest representations are selected with the same rules and are directly given by our extensive experiments. This process could be achieved automatically by inserting algorithms of corresponding rules. However, the choice in Figure 5 is manually selected for the time-saving purpose.

It seems better to choose a locally optimal k around the number of core concepts, so as to restrict the number of clusters within a suitable range for ontology learning purpose. This assumption takes the characteristics of primitive concepts into considerations. However, it rejects the possibilities of high-quality clustering along with smaller clusters. Therefore, in replace of the local optimization approach, global optimization of all indices is preferred to choose parameters of k-means clustering for every representation.

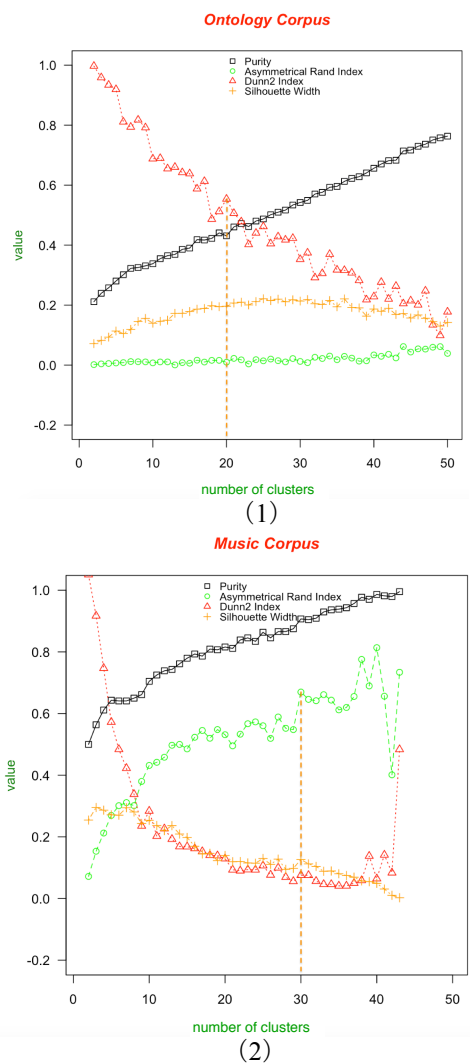

Figure 5: Example of Parameter Selection with K-Means

## 6.3 Interpretation of Clustering

Term clustering of feature representations is expected to capture core concepts labels in relevant with its syntactic context. Apart from the two distinct methods of feature extraction, such as NPs-VPCs couples and word embedding techniques, the influence of encoding techniques also changes the clustering quality.

Table 4 indicates the evaluation of clustering with golden standards. Encoding representations of NPs are denoted with corresponding techniques, such as 'NP_VPC_tfidf' and 'NP_VPC_NMF'. While the word embedding representation is said as 'NP_w2v'.

Table 4: K-Means and Affinity Propagation Evaluation

| | Corpus | Feature Representation | # optimal cluster/#core concepts | Purity | Asymm Rand Index | Dun2 Index | Silhouette Width |
|---|---|---|---|---|---|---|---|
| KM | Ontology | NP_VPC | 2.5(20/8) | 0.430 | 0.009 | 0.553 | 0.196 |
| | | NP_VPC_tfidf | 1.875(15/8) | 0.391 | 0.007 | **0.927** | **0.394** |
| | | NP_VPC_NMF | 3.375(27/8) | **0.545** | 0.067 | 0.381 | 0.390 |
| | | NP_w2v | 2.25(18/8) | 0.535 | **0.149** | 0.648 | 0.147 |
| | Music | NP_VPC | 6(30/5) | **0.906** | 0.6696 | 0.075 | 0.126 |
| | | NP_VPC_tfidf | 2(10/5) | 0.060 | 0.085 | 0 | 0 |
| | | NP_VPC_NMF | 3.4(17/5) | 0.761 | 0.348 | 0.559 | **0.292** |
| | | NP_w2v | 4(20/5) | 0.871 | **0.736** | 0.688 | 0.135 |
| AP | Ontology | NP_VPC | 1.87(15/8) | 0.450 | 0.044 | 0.730 | -0.131 |
| | | NP_VPC_tfidf | 1.75(14/8) | 0.410 | 0 | 0.738 | **-0.078** |
| | | NP_VPC_NMF | 2.625(21/8) | **0.492** | 0.096 | 0.743 | -0.108 |
| | | NP_w2v | 1.5(12/8) | 0.445 | 0.048 | **0.789** | -0.099 |
| | Music | NP_VPC | 0.6(3/5) | 0.666 | -0.11 | **0.921** | -0.02 |
| | | NP_VPC_tfidf | 0.4(2/5) | **0.75** | 0 | 1 | **0** |
| | | NP_VPC_NMF | 0.8(4/5) | 0.666 | -0.66 | 0.860 | -0.03 |
| | | NP_w2v | 2.4(12/5) | 0.445 | **0.048** | 0.789 | -0.09 |

In the upper half part of Table 4, diverse feature representations are clustered with k-means. In the aspect of the corpus, it is evident that Music corpus gives higher purity and higher Asymmetric Rand Index than that of Ontology Learning corpus. It can be due to that bigger corpus (Music Corpus) provides significant contextual features to cluster terms with taxonomic relations. On contrary, poor Dunn2 Index and Silhouette width in Music Corpus indicate terms do not compact closely with others, implying that features in bigger corpus have less similarity to others.

In the bottom part of Table 4, terms are clustered with affinity propagation algorithm. It is noticeable that NP_VPC representation family outperforms word2vec representation in Music Corpus. For NP_VPC representation family, the number of clusters in Music corpus is rather lower than that in Ontology Learning Corpus and the purity stays in a higher level. It is reasonable to infer that NP_VPC family are well suitable for AP clustering algorithm.

Comparing these two clustering methods, AP has a higher Dunn2 Index than that of K-Means, which means clusters of AP compact well. However, the negative silhouette width with AP indicates the intersection of clusters, which means feature similar terms probably share different labels. That is inevitable in linguistic because the similar context of terms could not straightly infer to the same meaning of them.

In terms of the encoding representations, Tf-Idf representations provide unevenly lower accuracy and higher compactness in clusters. While NMF representations have a good clustering quality overall.

In general, NP_VPC family representations appear to have a better clustering quality with AP clustering method than word2vec representation. On the other hand, encoding representations show an enhanced quality of clustering with K-Means. Precisely, NMF representations are prominent in most clustering situations.

## 7 CONCLUSIONS

Many works suggest making use of core ontology to build the modular ontology. However, most of these efforts are manually constructed and seldom in automatic approach. Term clustering according to a core ontology supports modular ontology construction without artificial demands. Taxonomic relations are constructed by gathering of NPs appearing with prominent syntactic roles after VPCs respecting to *core concepts*. Successfully we constructed feature space with these characteristics from two specialized corpora. To tackle the problem of sparsity, we benefit from feature selection and feature extraction techniques, such as adjusted Tf-Idf algorithm and NMF technique. Apart from that, word2vec is also compared as a benchmark. Along with all the extended representations, terms are clustered by K-Means and affinity propagation algorithm. We found that co-occurrence feature space appearing with syntactic roles, is proved to have a better clustering quality with the affinity propagation algorithm than that of K-Means. Furthermore, the usage of NMF on co-occurrence matrix could prominently improve clustering performance.

From the comparison of term clustering frameworks, we recommend beginning with a bigger domain-specific corpora. Since the syntactic relations between noun phrases and verbs are insufficient as features representation, with the assistance of encoding techniques, it gives rather convincing results in term clustering, which provides us a guideline for modular ontology building.

In the future work, we would like to benefit from prior knowledge of core concepts to assist in term selection process, so as to consider the characteristics of terms that related to core ontology. Furthermore, the morphological analysis could also help to merge specific terms into a general concept, which gives more distinguishing features of term clustering.

# REFERENCES

Aggarwal, C.C. and Zhai, C., 2012. A survey of text clustering algorithms. In *Mining text data* (pp. 77-128). Springer, Boston, MA.

Borgo, S. and Leitão, P., 2004, October. The role of foundational ontologies in manufacturing domain applications. In *OTM Confederated International Conferences" On the Move to Meaningful Internet Systems"* (pp. 670-688). Springer, Berlin, Heidelberg.

Buitelaar, P., Olejnik, D. and Sintek, M., 2004, May. A protégé plug-in for ontology extraction from text based on linguistic analysis. In *European Semantic Web Symposium* (pp. 31-44). Springer, Berlin, Heidelberg.

Burita, L., Gardavsky, P. and Vejlupek, T., 2012. K-GATE Ontology Driven Knowledge Based System for Decision Support. *Journal of Systems Integration*, *3*(1), p.19.

Cimiano, P., de Mantaras, R.L. and Saitia, L., 2004. Comparing conceptual, divisive and agglomerative clustering for learning taxonomies from text. In *16th European Conference on Artificial Intelligence Conference Proceedings* (Vol. 110, p. 435).

Competitions.codalab.org. (2018). *CodaLab - Competition*. [online] Available at: https://competitions.codalab.org/competitions/17119#learn_the_details-terms_and_conditions [Accessed 12 Jun. 2018].

Dunn, J.C., 1974. Well-separated clusters and optimal fuzzy partitions. *Journal of cybernetics*, *4*(1), pp.95-104.

Faure, D. and Nédellec, C., 1998. Asium: Learning subcategorization frames and restrictions of selection.

Fernández-López, M., Gómez-Pérez, A. and Juristo, N., 1997. Methontology: from ontological art towards ontological engineering.

Frey, B.J. and Dueck, D., 2007. Clustering by passing messages between data points. *science*, *315*(5814), pp.972-976.

Gábor, K., Zargayouna, H., Tellier, I., Buscaldi, D. and Charnois, T., 2016, October. Unsupervised Relation Extraction in Specialized Corpora Using Sequence Mining. In *International Symposium on Intelligent Data Analysis* (pp. 237-248). Springer, Cham.

Gamallo, P. and Bordag, S., 2011. Is singular value decomposition useful for word similarity extraction?. *Language resources and evaluation*, *45*(2), pp.95-119.

Gillis, N., 2014. The why and how of nonnegative matrix factorization. *Regularization, Optimization, Kernels, and Support Vector Machines*, *12*(257).

Gruber, T.R., 1993. A translation approach to portable ontology specifications. *Knowledge acquisition*, *5*(2), pp.199-220.

Harris, Z., 1954. Distributional structure.(J. Katz, Ed.) Word Journal Of The International Linguistic Association, 10 (23), 146-162.

Hartigan, J.A. and Wong, M.A., 1979. Algorithm AS 136: A k-means clustering algorithm. *Journal of the Royal Statistical Society. Series C (Applied Statistics)*, *28*(1), pp.100-108.

Hubert, L. and Arabie, P., 1985. Comparing partitions. *Journal of classification*, *2*(1), pp.193-218.

Kenneth, B. and Akitaka, M. (2018). [online] Cran.r-project.org. Available at: https://cran.r-project.org/web/packages/spacyr/spacyr.pdf [Accessed 12 Jun. 2018].

Kutz, O. and Hois, J., 2012. Modularity in ontologies. *Applied Ontology*, 7(2), pp.109-112.

Lee, D.D. and Seung, H.S., 1999. Learning the parts of objects by non-negative matrix factorization. *Nature*, *401*(6755), p.788.

Mikolov, T., Chen, K., Corrado, G. and Dean, J., 2013. Efficient estimation of word representations in vector space. *arXiv preprint arXiv:1301.3781*.

Nlp.stanford.edu. (2018). *Evaluation of clustering*. [online] Available at: https://nlp.stanford.edu/IR-book/html/htmledition/evaluation-of-clustering-1.html

Oberle, D., Lamparter, S., Grimm, S., Vrandečić, D., Staab, S. and Gangemi, A., 2006. Towards ontologies for formalizing modularization and communication in large software systems. *Applied Ontology*, *1*(2), pp.163-202.

Pal, N.R. and Biswas, J., 1997. Cluster validation using graph theoretic concepts. *Pattern Recognition*, *30*(6), pp.847-857.

Rdrr.io. (2018). *silhouette: Compute or Extract Silhouette Information from Clustering* [online] Available at: https://rdrr.io/cran/cluster/man/silhouette.html [Accessed 6 Jun. 2018].

Smith, J., 1998. *The book*, The publishing company. London, 2[nd] edition.